\documentstyle[preprint,aps]{revtex}
\begin{document}
\draft

\title
{Fermi edge restoration in the Tomonaga-Luttinger model with impurities}
\author{C.M.Na\'on, M.C.von Reichenbach and M.L.Trobo\\
{\normalsize \it Depto. de F\'\i sica.  Universidad
Nacional de La Plata.}\\
{\normalsize \it CC 67, 1900 La Plata, Argentina.}\\
{\normalsize \it Consejo Nacional de Investigaciones
Cient\'\i ficas y T\'ecnicas, Argentina.}\\}

\date{October 1995}
\maketitle
\thispagestyle{headings}
\markright{thepage}
\begin{abstract}
We study the Tomonaga-Luttinger model in the presence of magnetic
(Kondo-like) impurities. By using a recently proposed field-theoretical
approach to non-local bosonization we obtain the effective action
describing the  low-energy charge and spin density fluctuations.
{}From this action the dispersion relations of the collective modes are
readily found. We also compute the momentum distribution and
show that the electron-impurity scattering allows to have restoration
of the Fermi liquid behavior.(Contact naon@venus.fisica.unlp.edu.ar).
\end{abstract}
\pagenumbering{arabic}

\vspace{5cm}

\newpage
\narrowtext
\hspace*{0.6cm} Recent advances in the field of nanofabrication have allowed to
build ultranarrow semiconductor structures in which the motion of the electrons
is confined to one dimension \cite{PT} . These developments have triggered an
intense experimental \cite{WH} and theoretical \cite{DS} activity  in the last
few years.\\
\hspace*{0.6cm} One of the main tools for the theoretical study of the
one-dimensional (1D) electron system is the Tomonaga-Luttinger model (TL)
\cite{To}-\cite{ML} which describes a non-relativistic gas of massless
particles
with linearized free dispersion relation and two-body interactions. This model
displays the so called Luttinger liquid (LL) behavior \cite{H} : the jump
discontinuity in the electron momentum distribution is washed out as soon as
the interaction is switched on, even at $T=0$. Therefore, based on this model
one would expect that the experimental results should indicate a clear
deviation
from the normal Fermi liquid behavior (FL). However, the available experimental
data are consistent with the existence of the edge singularity in the momentum
distribution \cite{PL} .This fact raises an interesting question concerning the
validity of the TL model as a description of real 1D systems.\\
\hspace*{0.6cm} The main purpose of this letter is to show how the TL model can
be easily modified in order to have a restoration of the Fermi edge
singularity.
To this end we shall use a recently proposed  path-integral approach to
non-local
bosonization \cite{NRT}. In this framework the usual TL action can be expressed
as certain limit of a non-local QFT in which non-locality allows to consider
general potentials responsible for the two-body interactions. What we do here
is to add a Kondo-like term to the usual electron-electron scattering term of
the TL model and show under which conditions the presence of a magnetic
impurity can lead to normal FL behavior. Although the role of impurities in
the reappearance of the Fermi surface has been previously examined in another
model \cite{Hu}, we think it is specially relevant to have the possibility of
such a mechanism within the context of a well established exactly soluble
system like the TL model.
\newpage
\hspace*{0.6cm} Following the lines of Ref \cite{NRT} we consider the action

\begin{eqnarray}
 S  & = & \int d^2x~  [  \bar{\Psi} i \raise.15ex\hbox{$/$}
 \kern-.57em\hbox{$\partial$} \Psi  -  \int  d^2y ~
 J_{\mu}^{a}(x)V_{(\mu)}^{a b}(x,y) J_{\mu}^{b}(y)  - \nonumber\\
 & - & \int  d^2y ~
 J_{\mu}^{a}(x)U_{(\mu)}^{a b}(x,y) S_{\mu}^{b}(y)
+ d^{\dagger} i \partial_{t} d  ],
\label{1}
\end{eqnarray}

\noindent where
\[ \Psi = \left( \begin{array}{c}
              \Psi_{1} \\
              \Psi_{2} \\
              \end{array} \right)  \]

\noindent is the electron field ($\Psi_1 (\Psi_2)$ is associated to right
(left) movers),
\[ d = \left( \begin{array}{c}
                            d_{1}  \\
                            d_{2}  \\
                            \end{array} \right) \]

\noindent represents an impurity field with vanishing kinetic energy
\cite{FS}, and
\[
J_{\mu}^{a} = \bar{\Psi} \gamma_{\mu} \lambda^{a} \Psi,
\]
\[
S_{\mu}^{a} = \bar d \gamma_{\mu} \lambda^{a} d ,
\]

\noindent with $\lambda^a (a =0,1,2,3)$ the U(2) generators related to charge
and spin conservation. As shown in \cite{NRT}, the action (\ref{1}) with
$U_{(\mu)} = 0$, describes the zero-temperature TL model when only forward
scattering diagrams are taken into account. (For simplicity we set $v_{F} =1 $
and $p_{F} = 0$. The dependence on the Fermi momentum $p_{F}$ will be consider
later. See eq.(\ref{17})). If we disregard spin-flipping processes we can
restrict our analysis to the maximal Abelian subgroup of U(2), generated by
$\lambda^0$ and $\lambda^1$. The two-body potential matrices are diagonal
whose elements can be written in terms of the g-functions defined by S\'olyom
\cite{So} as

\begin{eqnarray*}
V_{(0)}^{0 0} &  = & \frac{1}{4} ( g_{4\parallel} + g_{4\perp} + g_{2\parallel}
+ g_{2\perp}), \\
V_{(0)}^{11} & = & \frac{1}{4} ( g_{4\parallel} - g_{4\perp} + g_{2\parallel}
- g_{2\perp} ),\\
V_{(1)}^{0 0}  & = & \frac{1}{4} (- g_{4\parallel} - g_{4\perp} +
g_{2\parallel}
+ g_{2\perp}), \\
V_{(1)}^{11}&  = & \frac{1}{4} (- g_{4\parallel} + g_{4\perp} + g_{2\parallel}
- g_{2\perp} ).
\end{eqnarray*}

The TL model, with charge-density fluctuations only, corresponds to
$V_{(0)}^{11} = V_{(1)}^{11} = 0 $. In a completely analogous way we
introduce the potentials that couple electron and impurity currents
in the form
\begin{eqnarray*}
U_{(0)}^{0 0} & = & \frac{1}{4} ( h_{4\parallel} + h_{4\perp} + h_{2\parallel}
+ h_{2\perp}), \\
U_{(0)}^{11} & = & \frac{1}{4} ( h_{4\parallel} - h_{4\perp} + h_{2\parallel}
- h_{2\perp}) , \\
U_{(1)}^{0 0} &  = & \frac{1}{4} (- h_{4\parallel} - h_{4\perp} +
h_{2\parallel}
+ h_{2\perp}),  \\
U_{(1)}^{11} & = & \frac{1}{4} (- h_{4\parallel} + h_{4\perp} + h_{2\parallel}
- h_{2\perp} ).
\end{eqnarray*}
This description includes both charge and spin density interactions. The
Kondo interaction, i.e. the coupling between spin densities only, corresponds
to the case $U_{(0)}^{00} = U_{(1)}^{00} = 0$.\\
\hspace*{0.6cm} Let us now consider the partition function of the system
\begin{equation}
Z = \int D\bar{\Psi} D\Psi D \bar d  Dd Da~  exp - [S + \int d^2x~ a (d^{\dag}d
- n)],
\label{5}
\end{equation}
where $S$ is given by (\ref{1}) and $a$ is a Lagrange multiplier that fixes
the impurity number to n. (Please see \cite{FS} for details concerning the
treatment of the impurity). Exactly as we did in \cite{NRT} for the ``clean"
system ($U_{(\mu)} =0$), we express (\ref{5}) in terms of fermionic
determinants.
This can be achieved by introducing auxiliary vector fields $A_{\mu}, B_{\mu}$.
After some standard manipulations we get
\begin{equation}
Z =  \int DA DB e^{-S'[A,B]} det ( i\raise.15ex\hbox{$/$}\kern-.57em
\hbox{$\partial$} - \sqrt{2} \not \!\! A ) det ( i \gamma_{0} \partial_{t}
-  \not \!\! B ),
\label{6}
\end{equation}

\noindent with
\begin{eqnarray}
S'[A,B] & = & \int d^2x ~ d^2y [ \sqrt{2} B_{\mu}^{a}(x) a_{(\mu)}^{a b}(x,y)
A_{\mu}^{b}(y)  -  \nonumber \\ & - & B_{\mu}^{a}(x) \int d^2u ~ d^2v
{}~ a_{(\mu)}^{a c}(x,u) ( b_{(\mu)}^{-1})^{c d}(u,v) a_{(\mu)}^{d b}(v,y)
B_{\mu}^{b}(y)], \nonumber\\
\label{7}
\end{eqnarray}

\noindent where we have defined the inverse potentials $a_{(\mu)}$ and
$b_{(\mu)}$ through the identities

\begin{eqnarray}
\int d^2u V_{(\mu)}^{ac}(x,u) b_{(\mu)}^{cd}(u,y) & = & \delta^{ad} \delta^{2}
(x-y),
\nonumber \\
\int d^2u U_{(\mu)}^{ac}(x,u) a_{(\mu)}^{cd}(u,y) & = & \delta^{ad} \delta^{2}
(x-y).
\label{8}
\end{eqnarray}
The fermionic determinant involving the field $A_{\mu}$ can be
readily computed by well-known path-integral techniques \cite{GS} based on the
change

\begin{eqnarray}
\Psi & = & e^{- \sqrt{2} (\gamma_{5} \Phi + i \eta)} \chi,      \nonumber \\
\bar{\Psi} & = &\bar{\chi} e^{- \sqrt{2}(\gamma_{5} \Phi - i \eta)},
\nonumber\\
A_{\mu} & = & \epsilon_{\mu \nu} \partial_{\nu} \Phi + \partial_{\mu} \eta,
\label{9}
\end{eqnarray}
with $\Phi = \Phi^{a} \lambda^{a} , \eta = \eta^{a} \lambda^{a}$ .

The result is
\[det ( i\raise.15ex\hbox{$/$}\kern-.57em\hbox{$\partial$} - \sqrt{2}
\not \!\! A ) = (det i\raise.15ex\hbox{$/$}\kern-.57em\hbox{$\partial$})
exp \frac{1}{\pi} \int d^2x ~\Phi \Box  \Phi. \]
 \hspace*{0.6cm} Concerning the determinant associated with the impurity, its
 evaluation is more subtle. However, we were able to extend the standard
 technique \cite{GS} to solve this particular problem. (The details of this
 calculation will be reported elsewhere. An alternative treatment can be
 found in \cite{FS}). We thus get

 \begin{eqnarray*}
 det( i \gamma_{0} \partial_{t}-\not \!\! B )&=&det( i\gamma_{0}\partial_{t})
 exp  \frac{-1}{4 \pi} \int d^2x \{B_{0}^{2} - B_{1}^{2}  + \\
& + & \int d^2y \Theta ( x_{0} - y_{0}) \partial_{x} \delta (x_{1} - y_{1})
[ B_{0}(x) B_{1}(y) + B_{1}(x) B_{0}(y) ] \} .
 \end{eqnarray*}

 Putting all this together and going to momentum space one gets a bosonic
 action depending on $\Phi$ and $\eta$ (the collective modes) and the
 impurity variables $B_{0}$ and $B_{1}$.  These last fields, in turn, can
 be easily integrated out. Thus, we finally obtain
 \begin{equation}
 Z = \int D\Phi D\eta~exp - \left\{ S_{eff}^{00} + S_{eff}^{11} \right\},
 \label{12}
 \end{equation}
\noindent with
 \begin{eqnarray}
 S_{eff}^{ii} & = & \frac{1}{(2 \pi)^{2}} \int d^2p
[ \Phi^{i}(p) A^{ii}(p) \Phi^{i}(-p) +
\eta^{i}(p) B^{ii}(p) \eta^{i}(-p)  + \nonumber \\
& + & \Phi^{i}(p) \frac{C^{ii}(p)}{2} \eta^{i}(-p) +
\eta^{i}(p) \frac{C^{ii}(p)}{2} \Phi^{i}(-p) ],
\label{13}
\end{eqnarray}

\noindent where
\begin{eqnarray}
 A(p) & = & \frac{1}{\Delta(p)}\left\{ \frac{p^{2}}{\pi} \Delta - a_{0} a_{1}
 \frac{p_{1}^{2}}{\pi} + \frac{1}{2 \pi} (a_{0}^{2} p_{1}^{2} - a_{1}^{2}
 p_{0}^{2}) -2 a_{0}^{2} a_{1}^{2} (\frac{p_{1}^{2}}{ b_{1}} + \frac{p_{0}^{2}}
 {b_{0}}) \right\}, \nonumber \\
B(p) & = & \frac{1}{\Delta(p)}\left\{ \frac{p_{1}^{2}}{\pi} a_{0} a_{1} +
\frac{1}{2 \pi} (a_{0}^{2} p_{0}^{2} - a_{1}^{2} p_{1}^{2}) - 2 a_{0}^{2}
a_{1}^{2} ( \frac{p_{1}^{2}}{b_{0}} + \frac{p_{0}^{2}}{b_{1}}) \right\},
\nonumber \\
C(p) & = & \frac{1}{\Delta(p)}\left\{ \frac{ a_{0} a_{1}}{\pi} ( \frac{p_{1}^
{3}}{p_{0}} - p_{0} p_{1} ) + \frac{ p_{0} p_{1}}{\pi} (a_{0}^{2} + a_{1}^{2})
+
4 p_{0} p_{1} a_{0}^{2} a_{1}^{2} (\frac{1}{ b_{0}} - \frac{1}{b_{1}})\right\},
\nonumber \\
\Delta (p)& =& \frac{ p_{1}^{2}}{ 4 \pi^{2} p_{0}^{2}} + 4 ( \frac{1}{ 4 \pi} -
\frac{ a_{1}^{2}}{ b_{1}}) ( \frac{1}{ 4 \pi } + \frac{a_{0}^{2}}{ b_{0}}).
\nonumber
\end{eqnarray}

\hspace*{0.6cm} For the sake of clarity we have omitted $ii$ superindices in
the above expressions, which are written in terms of the Fourier transforms
of the inverse potentials defined in (\ref{8}). (Note that $b_{\mu}(p)
= V_{(\mu)}^{-1}(p)$ and $a_{\mu}(p) = U_{(\mu)}^{-1}(p)$ ).\\
\hspace*{0.6cm} Eqs (\ref{12}) and (\ref{13}) are our first non-trivial
results. Indeed, we have obtained a completely bosonized action describing
the dynamics of charge density (CDW) and spin density (SDW) excitations
(associated with the fields $\Phi^{0}, \eta^{0}$ and $\Phi^{1}, \eta^{1}$
respectively). As we can see from (\ref{12}), these modes remain decoupled
as in the impurity free case. The dispersion relations are given by

 \begin{equation}
\omega^{2}(q) = q^{2} \frac{ 1 + \frac{2}{\pi} V_{(0)} + \frac{1}{2 \pi ^{2}}
(U_{(0)}^{2} - 2 U_{(0)} U_{(1)})}{1 + \frac{2}{\pi} V_{(1)} -
\frac{1}{2 \pi ^{2}} U_{(1)}^{2}},
 \label{14}
 \end{equation}

where, once again, we have omitted $00$ ($11$) superindices
in the potentials, corresponding to CDW's (SDW's). We have also gone back
to real frecuencies: $p_{0} = i \omega$, $p_{1} = q$.\\
 \hspace*{0.6cm} From now on we shall specialize the discussion to the TL model
 with a Kondo-like interaction. Thus, we set
 \begin{eqnarray*}
 V_{(0)}^{11} & = & V_{(1)}^{11} = 0 = U_{(0)}^{00} = U_{(1)}^{00}, \\
 \frac{2}{\pi} V_{(0)}^{00}& = & r, \\
 \frac{2}{\pi} V_{(1)}^{00} & = & s, \\
 (U_{(0)}^{11})^{2} & = & (U_{(1)}^{11})^{2}  =  2 \pi^{2} t.
\end{eqnarray*}

For repulsive electron-electron interaction one has $r>0$ and $s>0$, whereas
$t>0$ for both ferromagnetic and antiferromagnetic Kondo coupling. In order to
study the momentum distribution we compute the fermionic 2-point
 function. To be specific we consider $G_{1 \uparrow}$ (similar expressions are
 obtained for $G_{2 \uparrow}$, $G_{1 \downarrow}$ and $G_{2 \downarrow}$ ),
  \begin{eqnarray}
  G_{1 \uparrow}(x,y) & = & <\Psi_{1 \uparrow} (x) \Psi^{\dag}_{1 \uparrow}(y)>
\nonumber \\
  & = & G_{1 \uparrow}^{(0)} (x,y) <e^{\sqrt{2}\left\{[ \Phi^{0}(y)
  -\Phi^{0}(x)] +
i[\eta^{0}(y) - \eta^{0}(x)]\right\}}>_{00}  \times  \nonumber \\
& \times & <e^{\sqrt{2}\left\{[ \Phi^{1}(y) -\Phi^{1}(x)] +
i[\eta^{1}(y) - \eta^{1}(x)]\right\}}  >_{11},
  \label{16}
  \end{eqnarray}
where we have used (\ref{9}) to relate $G_{1 \uparrow}$ with the free
propagator $G_{1 \uparrow}^{(0)}$ , in which the Fermi momentum $p_{F}$ is
easily incorporated \cite{NRT}. The symbol $ <>_{ii}$ means v.e.v. with
respect to the action (\ref{13}). Taking the limit $ z_{0} \rightarrow 0$
($z = x - y$ ) in (\ref{16}) and inserting the result in the definition of
the momentum distribution $N_{1 \uparrow}(q)$, one gets
 \begin{equation}
 N_{1 \uparrow}(q) = C(\Lambda) \int dz_{1} \frac{e^{-i(q - p_{F})z_{1}}}
 {z_{1}} e^{-\int dp_{1} \frac{ 1 - cos p_1 z_1 }{p_1}  \Gamma (p_1)},
 \label{17}
 \end{equation}
 where $C(\Lambda)$ is a normalization constant depending on an ultraviolet
cutoff $\Lambda$, and $\Gamma(p_1)$ depends on $p_1$ through the potentials
in the form
\begin{equation}
\Gamma (r,s,t) = f(r,0) + f(s, t) - h(s,t).
\label{18}
\end{equation}
In the above equation we have defined the functions
\begin{equation}
f(s,t) = \frac{( \mid 1 + s - t \mid^{1/2} - \mid 1 -t \mid^{1/2})^{2}}
{\mid 1 + s - t \mid^{1/2} \mid 1 - t \mid^{1/2}},
\label{19}
\end{equation}
\begin{equation}
h(s,t)= \frac{ 2 t ( t - t_{c} )}{  \mid 1 + s - t \mid^{3/2}
\mid 1 - t \mid^{1/2}},
\label{20}
\end{equation}
with $t_{c} = 9/8 - 1/2 ( s - 1/2 )^{2}$.\\
\hspace*{0.6cm} We have now reached the main point of our discussion. We want
to determine under which conditions it is possible to have $\Gamma(r,s,t) = 0$,
since in this case one immediately has $N_{1 \uparrow}(q) \propto \Theta(q -
p_{F})$, i.e. normal FL behavior. First of all we note that for $t = 0$
(impurity free case) $\Gamma (r,s,0)$ cannot vanish for any value of r and s
other than $r = s = 0$, which corresponds to the non-interacting Fermi gas.
On the other hand, the first two terms in (\ref{18}) are positive definite.
This means that the only chance to obtain $\Gamma = 0$, in a non-trivial
(interacting) situation, is to have $t>t_{c}$. This necessary condition
determines a ``critical" parabola in the space of potentials, $t(s) = t_{c}$,
below which FL behavior is forbidden. Enforcing also the condition  $\omega^{2}
> 0$ in (\ref{14}), in order to have normal modes with real frequencies,  one
finds two disjoint regions where the  FL edge could be, in principle, restored:
$t > 1 + s$  and $t < 1 , t > t_{c}$ . However, one also has
to satisfy $f(r,0) > 0$, which for $\Gamma = 0$ yields
\begin{equation}
F(s,t) = h(s,t) - f(s,t) > 0.
\label{21}
\end{equation}
\noindent A simple numerical analysis of $F(s,t)$ shows that the above
inequality is not fulfilled for $0 < t <1$. The electron-impurity coupling is
not strong enough in this region as to eliminate the LL behavior. On the
contrary, for $t > 1 + s$ equation (\ref{21}) can be always satisfied.
Moreover, in this region, for $\Gamma = 0$ we obtain the following analytical
solution for $r$ in terms of $F(s,t)$:
\begin{equation}
r = F^{2}/2 + 2 F + ( 1 + F/2) ( F^{2} + 4 F)^{1/2}.
\label{22}
\end{equation}

To illustrate this result it is useful to consider the local case,
corresponding
to contact interactions ($r,s$ and $t$ are constants). From the precedent
discussion we conclude that for $0 < t < 1$ (region I) one necesarilly has LL
behavior, whereas for $ t > s + 1$ (region III) FL behavior is admitted.
(See Fig. 1). Note that for $1 < t < s + 1$ (region II) the frecuency of the
SDW's becomes imaginary. Let us stress that, for $t > s + 1$, Eq(\ref{22})
defines a surface in the space of potentials on which FL behavior takes place.
This is our main result. One particular solution belonging to this ``FL
surface''
is obtained by choosing $s = 0$ in (\ref{22}), which yields

\[ r(t) = \frac{2t(3t-2) + 2(2t-1)\sqrt{3t^2 - 2t}}{(t-1)^2}.\]

This curve corresponds to the case in which the SDW's dispersion relation
is given by $\omega^2 = q^2$. For $t$ large $r$ approaches a minimum
value $r_{min}= 6 + 4\sqrt{3}$, a feature that is shared with each curve
on the FL surface. \\
\hspace*{0.6cm} In summary, we have studied a simple modification of
the TL model that describes the interaction of electrons with localized
magnetic impurities at zero temperature. By using a recently proposed
field-theoretical path-integral approach that enables us to treat non-local
interactions \cite{NRT}, we were able to obtain a completely bosonized
effective action that governs the dynamics of the collective modes of our
model (Eqs. (\ref{13}), (\ref{14})). This action, in turn, allowed us to
compute the electron momentum distribution (Eq.(\ref{17})). Finally, by
analyzing this result we have found one region in the space of potentials in
which the FL behavior can be recovered as an effect of the interaction between
the electrons and the magnetic impurities. Thus, we have presented a
qualitative
scheme that could be useful in the understanding of highly correlated 1D
systems,
in terms of the well-known TL model.
\vspace{1cm}

\section*{Acknowledgements}
The authors are partially supported by Fundaci\'on Antorchas (Buenos Aires,
Argentina) under grant A-13218/1-000069.

\begin{figure}
\caption{Three regions in the first quadrant of the s,t plane. In I
FL behavior is forbidden. In III the Fermi edge is restored for r(s,t)
given by Eq.(16). In II SDW's are absent.}
\label{Fig.1}
\end{figure}

\end{document}